\pdfoutput=1 
\documentclass{JINST}
\let\ifpdf\relax

\title{Operation and performance of the CMS tracker}

\author{Viktor Veszpremi for the CMS Collaboration$^a$\thanks{Supported by the Hungarian Scientific Research Fund under contract number OTKA NK81447.}\\
\llap{$^a$}Wigner Research Centre for Physics,\\
  1525 Budapest, P.O.Box 49, Hungary\\
E-mail: \email{veszpremi.viktor@wigner.mta.hu}}

\abstract{
The CMS silicon tracker consists of two tracking devices utilizing semiconductor technology: the inner pixel and the outer 
strip detectors. They operate in a high-occupancy and high-radiation environment presented by particle collisions 
in the LHC. The tracker detectors occupy the region around the center of CMS, where the LHC beams collide, between 
4 cm and 110 cm in radius and up to 280 cm along the beam axis. The pixel detector consists of 66 million pixels, covering 
about 1 m$^2$ total area. It is surrounded by the strip tracker with 10 million read-out channels covering about 200 m$^2$ total area. 
The proceedings describe the operational experience collected during the first three years of LHC running. 
Results include operational challenges encountered during data taking that influence the active fraction and 
read-out efficiency of the detectors. Details are given about the performance of the tracker at high occupancy with respect 
to local observables such as signal to noise ratio and hit reconstruction efficiency. Studies of radiation effects are presented 
with respect to the evolution of sensor bias, read-out thresholds in the inner pixels, and leakage current.
}

\keywords{LHC; CMS; Silicon Tracker}

\begin{document}

\section{Introduction}\label{sec:Introduction}

\begin{figure}[tbp] 
\centering
\includegraphics[height=5cm]{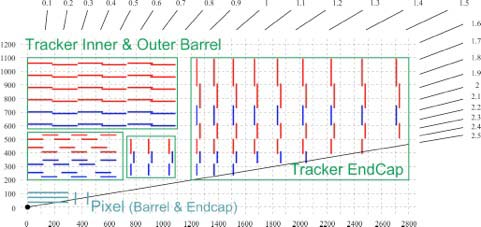}
\caption{Quarter of the $r-z$ slice of the CMS tracker, where the center of the tracker is at the left-bottom corner
of the drawing, the horizontal axis, to which the detector has a cylindrical symmetry, points along $z$, and the vertical 
axis points along the radius $r$. The LHC beams are parallel to the $z$ axis. Various pseudorapidity values are 
shown at the ends of the black lines. Tracker barrel and endcap modules in red are single-sided, those in blue are stereo.
}
\label{fig:CMSTracker}
\end{figure}

The CMS tracker is an all-silicon detector \cite{CMSExperiment} with a sensitive area 
of over 200 m$^2$. The sensors are arranged in concentric cylinders around the interaction region of 
the LHC beams and are situated in a 3.8 Tesla magnetic field. The purpose of the detector is to provide 
high precision measurement points in three dimensions along the curved trajectories of charged
particles up to pseudorapidities $|\eta|<2.5$. The best tracking efficiency is achieved in the barrel region, $|\eta|<0.9$ 
(figure \ref{fig:CMSTracker}). The charged particle tracks are used to reconstruct the positions
of the primary interaction and secondary decay vertices. The tracker allows for rapid and precise measurements 
with temporal and spatial resolutions that fulfill the challenges posed by the high luminosity LHC
collisions, which occur at a frequency of 40 MHz. The high particle fluence induces radiation damage, 
which also presents a challenge for the operation and data-reconstruction in the inner layers of the tracker.

The CMS tracker is comprised of two sub-detectors with independent cooling, powering, and 
read-out schemes. The inner sub-detector, the pixel detector, has a surface area of 1.1 m$^2$. It is segmented into 
66 million $n+$ pixels of size 100 $\mu$m by 150 $\mu$m implanted into $n$-type bulk with thickness of 285 $\mu$m and $p$-type back side. 
The detector has three layers in the barrel region at radii of 4.3 cm, 7.2 cm, and 11 cm, respectively, and two disks on each side of 
the barrel (the endcap regions) at 34.5 cm and 46.5 cm from the interaction point. The pixel detector contains 
15840 read-out chips (ROC), each reading an array of 52 by 80 pixels. The ROCs are arranged 
into modules which transmit data via 1312 read-out links.

The sub-detector surrounding the pixels, the strip detector, is segmented into 9.6 million $p+$ strips which are implanted into 
$n$-type bulk with thickness of 320 $\mu$m (500 $\mu$m) in the inner (outer) layers or disks and $n$-type back side. 
The pitch of the strips varies from 80 $\mu$m to 205 $\mu$m. The detector has 10 tracking layers in the barrel 
region that span radii from 25 cm to 110 cm and along the $z$ axis up to 120 cm: 4 layers in the inner barrel 
(TIB) and 6 in the outer barrel (TOB). It also has 12 disks in the endcap region with radii up to 110 cm 
and in $z$ up to 280 cm: 3 inner disks (TID) inside and 9 endcap disks (TEC) outside the TOB as shown in figure \ref{fig:CMSTracker}. 
Four layers in the barrel and multiple layers in the endcap regions of the strip detector are equipped with 
stereo modules allowing for 2D measurement. These modules have two silicon sensors mounted back-to-back with 
their strips aligned at a 100 mrad relative angle. Both sub-detectors are read out via a chain of analog 
electronic and optical links which are able to transmit absolute pulse height. In the pixel detector,
the pixel coordinates are also transmitted. For the strips, all data-processing happens in off-detector
electronics.

\section{Operation of the tracker}\label{sec:Operation of the Tracker}

\begin{figure}[tbp] 
\centering
\includegraphics[width=9.5cm,height=4.5cm]{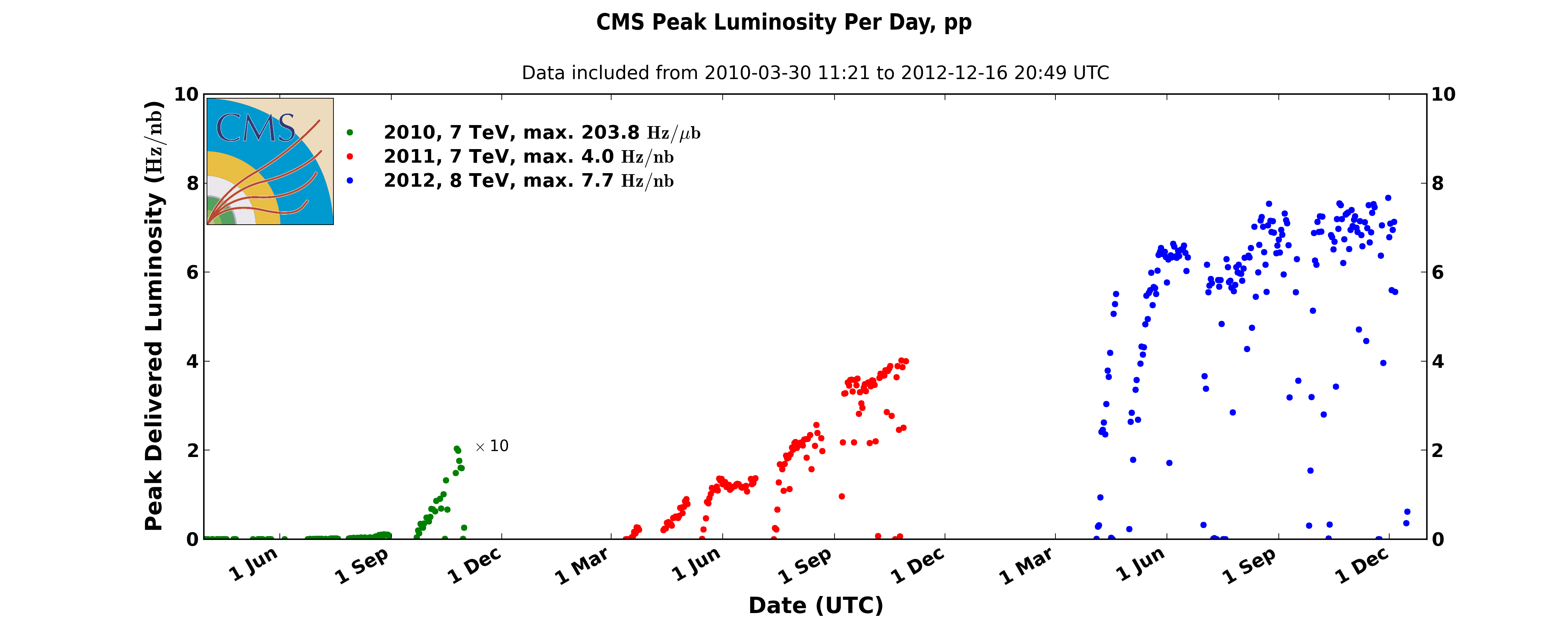}
\includegraphics[width=5.5cm,height=4.5cm]{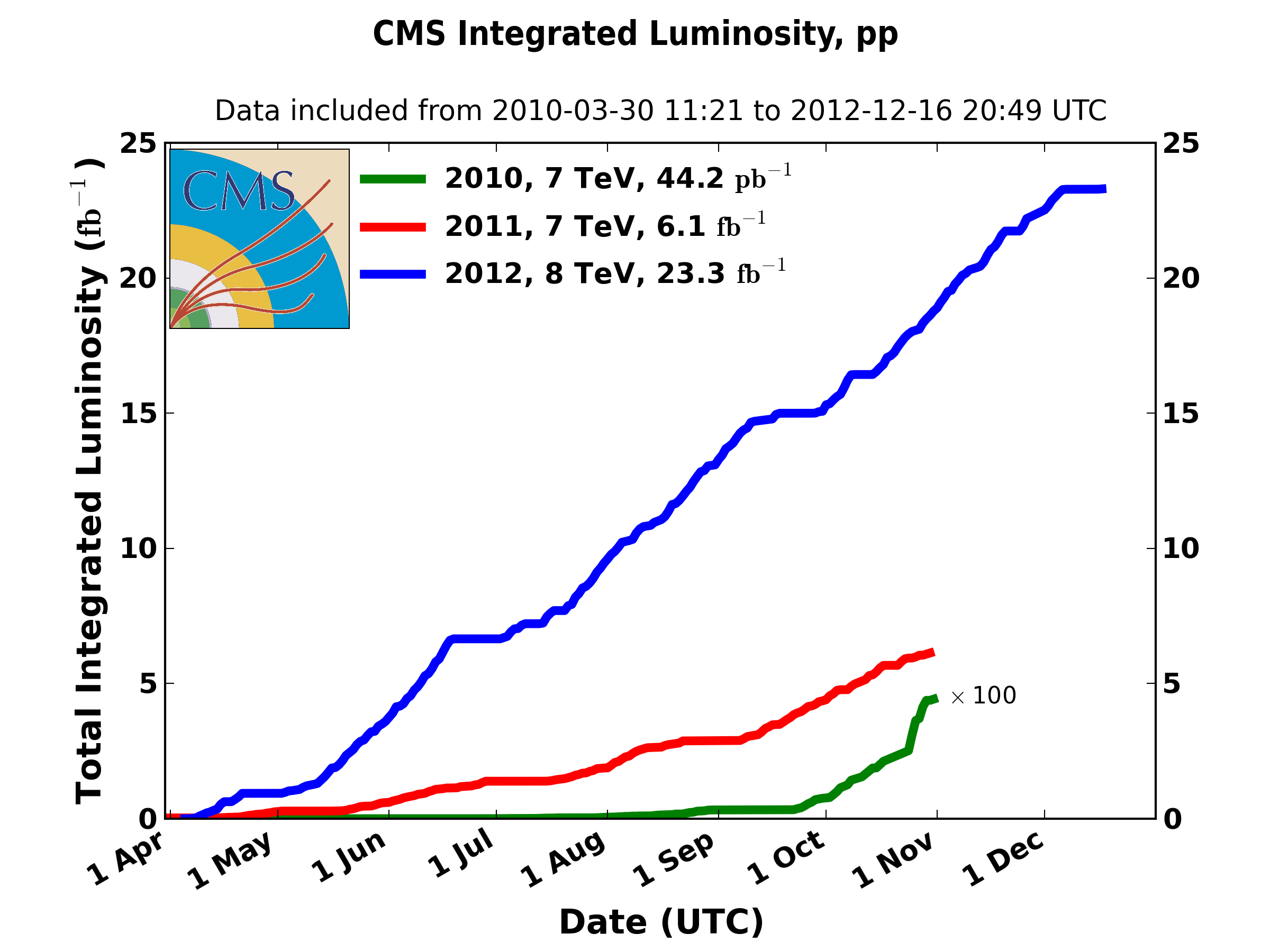}
\caption{LHC beam parameters. Peak (left) and cumulative (right) luminosities delivered to CMS during stable beams in p-p 
collisions as function of time. This is shown for 2010 (green), 2011 (red) and 2012 (blue) data-taking.\cite{CMSLumi}}
\label{fig:LHCBeamParameters}
\end{figure}

In the years of operation between 2010 and 2013, the LHC \cite{LHCMachine} has delivered about 6.1 fb$^{-1}$ integrated luminosity of 
proton-proton collision data at 7 TeV and about 23.3 fb$^{-1}$ at 8 TeV (figure \ref{fig:LHCBeamParameters}). CMS has recorded 
overall 93\% of these data \cite{CMSLumi}. 
The tracker was responsible for only about a third of the data lost, primarily because its high-voltage is ramped up only after stable 
collisions are declared. The reliability of the detector was constant over this period despite
the increasing challenge presented by the continuously increasing instantaneous luminosity. The LHC
reached its peak instantaneous luminosity of 7.7$\times$10$^{33}$cm$^{-2}$s$^{-1}$ in late 2012. High instantaneous
luminosity causes multiple proton-proton interactions in the detector, known as pile-up. The average pile-up in 2012 was 21 simultaneous 
proton-proton collisions, and half of all events had pile-up between 21 and 40. The impact of pile-up will be further discussed later.
By the time of the shutdown in 2013, about 2.3\% of the barrel and 7.2\% of the endcap modules of the pixel detector were inactive; 
mostly due to faulty wire-bonds or poor connections.
Over the same period of data-taking, about 2.5\% of the strip detector became inactive due to short-circuits in the control 
rings and HV lines, or as a result of faulty optical communications. Repair of
the damaged modules was part of the maintenance performed during 2013; up to 1.5\% of the pixel barrel, up to 0.5\% of the
pixel endcap, and up to 1\% of the strip detectors were deemed recoverable.

\section{Hit reconstruction}\label{sec:HitReconstruction}

\begin{figure}[tbp] 
\centering
\includegraphics[width=7cm,height=5cm]{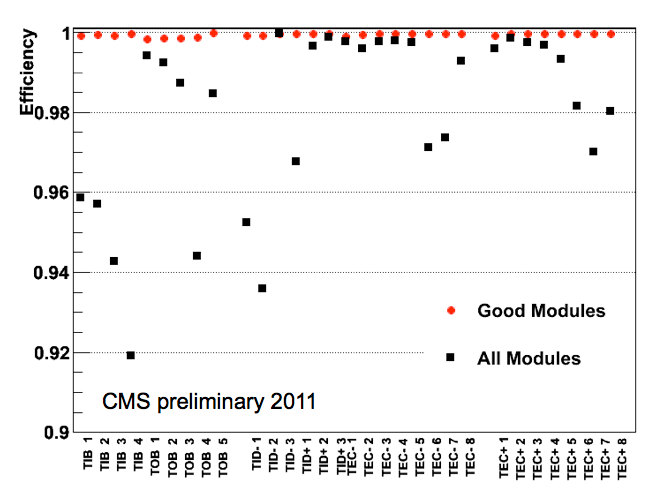}
\includegraphics[width=5cm,height=5cm]{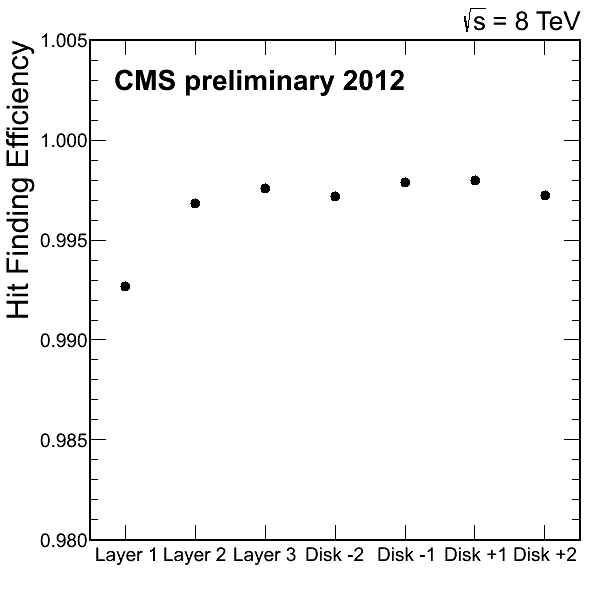}
\caption{Single hit efficiencies measured on the layers of the strip (left) and pixel (right) detectors. Only fully
operational modules were considered for the pixels \cite{TrackerDPG2013, CMSTrackingPerf2010}.}
\label{fig:HitEfficiency}
\end{figure}

The first step in processing the data prior to track reconstruction is the efficient detection 
of hits, which represent the positions in the tracker's sensors where charged particles passed through. 
Figure \ref{fig:HitEfficiency} shows the hit-finding efficiency in various
layers of the strip and pixel detectors. After full track reconstruction, the efficiency is measured as the 
fraction of the number of particles that are expected to pass through the fiducial regions of the sensors in a given layer 
for which corresponding hits are found. In the case of the strip detector, a hit is considered to be found if it is on
the same module in which the hit was expected to be observed. In the case of the pixel detector, a more stringent 
selection is required in order to address the higher particle occupancy: a hit must be found within a 500 $\mu$m radius of the expected 
intersection point. In both cases, the only particle trajectories that are considered are those which have reconstructed hits on 
adjacent layers. The layer efficiencies exceed 99\% in the strip detector and 99.5\% in the pixel detector 
with the exception of the innermost layer of the pixels (see explanation in section \ref{sec:LHCBeamEffectsDynamic}).

\begin{figure}[tbp] 
\centering
\includegraphics[width=5cm,height=5cm]{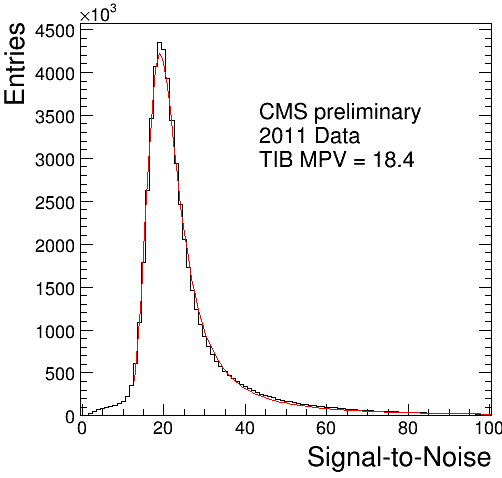}
\includegraphics[width=5cm,height=5cm]{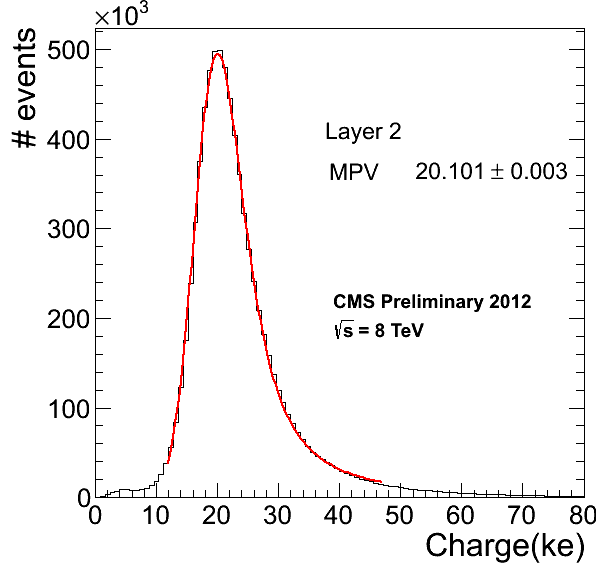}
\caption{Examples of a signal-to-noise ratio distribution in the inner barrel (TIB) of the strips (left) and total cluster charge distribution 
in the pixels (right) from hits on reconstructed particle tracks. Both most probable values (MPV) are corrected for the track impact angles \cite{TrackerDPG2013}.}
\label{fig:ClusterParameter}
\end{figure}

Besides the hit-finding efficiency, the resolution of the hit position is another important performance parameter of 
the tracker. They both rely on the complete measurement of clusters, contiguous groups of pixel or strip signals belonging to a hit, 
the formation of which depends on the path-length of the incident particles in the sensor and the Lorentz drift induced charge sharing. 
Figure \ref{fig:ClusterParameter} shows the signal-to-noise ratio in the strips and the cluster charge in the pixels for hits 
along reconstructed tracks. In both cases, the measurements are corrected for the impact angles of the incident particles,
and the yields are fitted with a Landau distribution convoluted with a Gaussian.

\section{LHC beam-induced effects}\label{sec:LHCBeamEffects}

The data-taking conditions of the tracker detectors operating in the LHC environment are determined by two factors: 
the instantaneous luminosity through the resulting particle occupancy in a series of collision events, and the 
integrated luminosity which is closely correlated with the accumulated irradiation dose encountered by the sensors. 
The effects of the former are short-term and follow dynamically the changing beam conditions. The channel occupancy
is defined as the number of pixel or strip measurements in an event divided by the number of all active pixels or strips.
It scales linearly with pile-up. At the average pile-up of 9, which was typical in 2011, the pixel occupancy was 2$\times$10$^{-4}$
in layer 1 and 0.5$\times$10$^{-4}$ in layer 3 of the barrel pixel. The strip occupancy was 9$\times$10$^{-3}$ in the innermost layer of the 
TIB and 10$^{-3}$ in the outermost layer of the TOB. The typical average pile-up increased by about a factor of 2.5 between 
2011 and the end of 2012. Radiation-induced effects are cumulative in time and are essentially irreversible. The 
radiation dose acquired between 2010 and 2012 was estimated to be about 40 kGy in the innermost and 10 kGy in the outermost 
layers of the pixel detector. The strip detector integrated about 5 kGy radiation in the innermost layer of the TIB and 
200 Gy in the outermost layer of the TOB. Both occupancy and radiation-induced effects are more important for the pixel 
detectors, as they are closer to the interaction region. In the following two subsections, studies performed on the pixel 
detector will be described.

\subsection{Short-term, dynamic effects}\label{sec:LHCBeamEffectsDynamic}

\begin{figure}[tbp] 
\centering
\includegraphics[width=5cm,height=5cm]{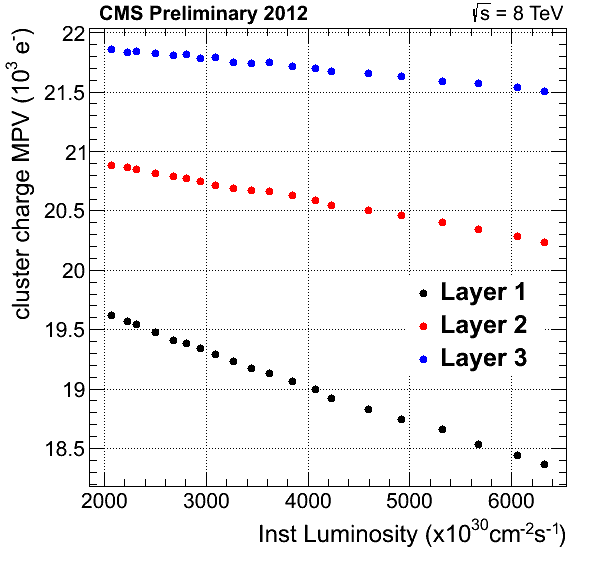}
\includegraphics[width=5cm,height=5cm]{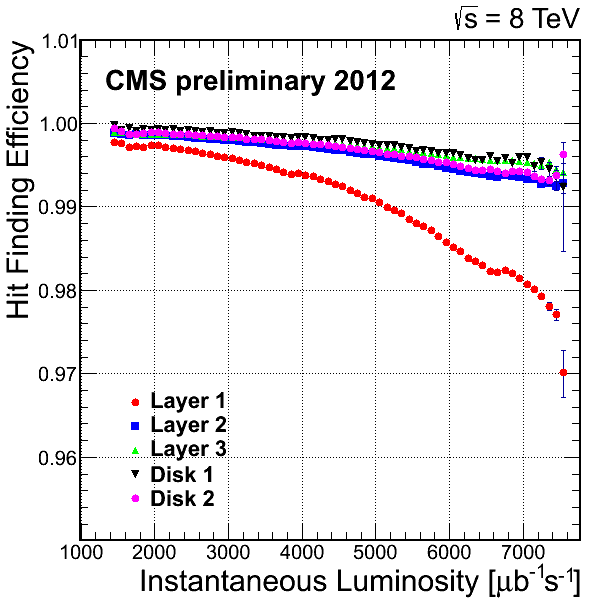}
\caption{Occupancy related effects on the read-out mechanisms of the ROCs. Both the charge calibration (left) and the
single hit efficiency (right) are functions of the instantaneous luminosity \cite{TrackerDPG2013}.}
\label{fig:OccupancyDependence}
\end{figure}

The power consumption of the ROCs increases in direct proportion to the particle occupancy in collision events,
leading to higher operating temperatures. The temperature dependence of the pixel charge calibration is considered to
be responsible for the decrease of the cluster charge at higher instantaneous luminosities, as shown in figure 
\ref{fig:OccupancyDependence} (left), where the most probable value of the Landau-fit on the cluster charge is plotted.

The CMS global trigger system needs about 4 $\mu$s to make a decision and send a signal to the pixel detector in order 
to indicate whether a collision event is to be read-out or discarded. The ROCs store their hit information 
in internal buffers during this time. At high occupancy, the ROC buffers may entirely fill up with older hits,
preventing the storage of hits in more recent events. Indirect evidence for the resulting loss of hit information can be 
seen in figure \ref{fig:OccupancyDependence} (right). It shows the dependence of the hit finding efficiency on the 
instantaneous luminosity, which is the most pronounced for the innermost layer, as expected. Direct study of this 
mechanism is not possible, therefore approximate loss rates can only be inferred from simulation.

Another example of short-term beam effects on data taking is the so-called Single Event Upset (SEU). Particles from 
collisions may randomly flip bits in the control registers of the ROCs and auxiliary electronics. SEUs may interrupt or degrade
data-taking. Their occurrence is proportional to the instantaneous luminosity and randomly distributed over time. A
reprogramming of the ROCs solves this problem, which may be automatically triggered by the read-out front end, or 
the data quality monitoring framework.

\subsection{Long-term, cumulative effects}\label{sec:LHCBeamEffectsCumulative}

\begin{figure}[tbp] 
\centering
\includegraphics[width=5cm,height=5cm]{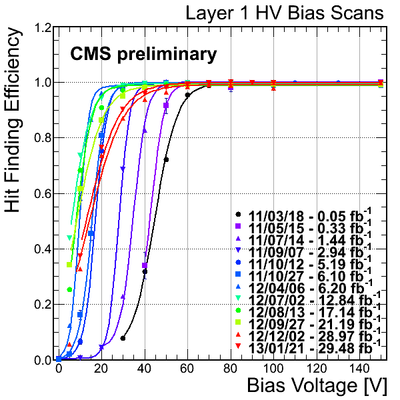}
\includegraphics[width=5cm,height=5cm]{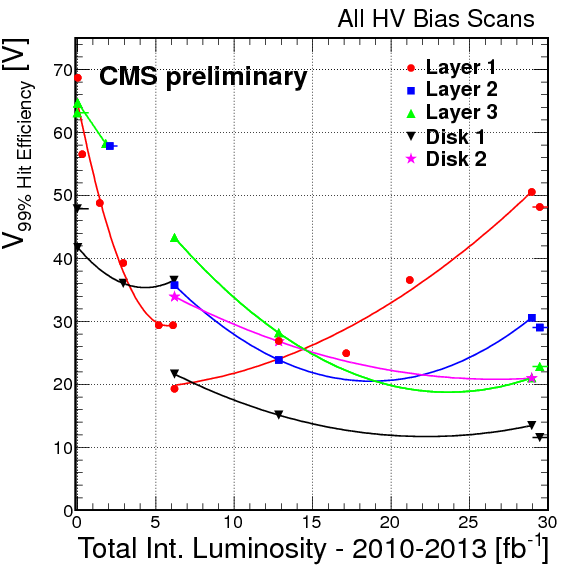}
\caption{Bias scans performed on the pixel detector (left) and the bias voltage corresponding to 99\% single hit efficiency (right) 
as function of the integrated luminosity \cite{TrackerDPG2013}. Multiple data points acquired in the same year are connected with 
quadratic fits in the right plot only in order to guide the eye, no underlying model is implied. }
\label{fig:BiasScan}
\end{figure}

Radiation induced damage in the silicon bulk was monitored throughout the LHC running between 2010 
and 2013. The bias voltage applied to the sensors during normal operation is 150 V in the barrel and 300 V in the endcap 
of the pixel detector. Special runs were performed regularly in which the bias voltage was increased in steps from zero to 
the operational voltages. The hit efficiency of the pixels was measured at each point, as shown in figure \ref{fig:BiasScan}.
The bias voltage that is needed to reach a depletion depth corresponding to full hit efficiency decreases with irradiation
at first, then increases as expected due to changes in the effective doping \cite{RossiFischerRoheWermes}. The bias voltage
at which the efficiency reaches 99\% on each layer is plotted as a function of the integrated luminosity. It 
shows evidence of space charge sign inversion in the first and second layers.

\begin{figure}[tbp] 
\centering
\includegraphics[width=7cm,height=5cm]{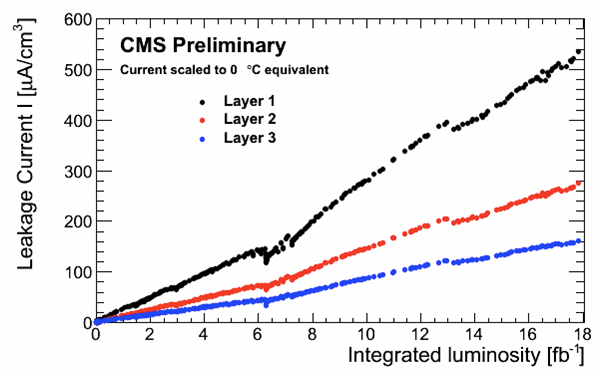}
\includegraphics[width=5cm,height=5cm]{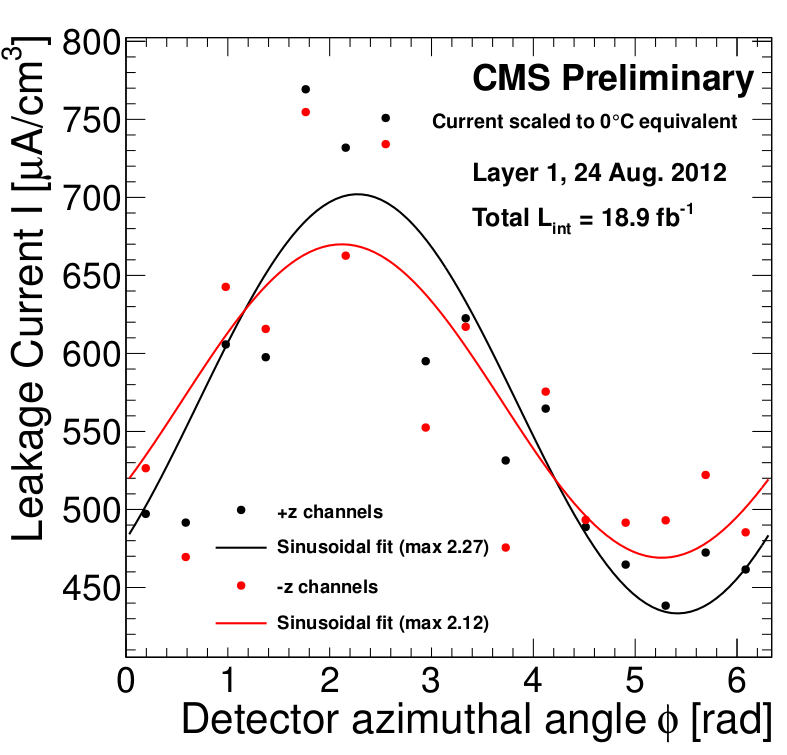}
\caption{Pixel leakage current scaled to $0\,^{\circ}\mathrm{C}$ operational temperature as a function of total irradiation (left) and azimuthal
angle (right) that circles around the symmetry axis of the detector. The uneven leakage current in the azimuthal angle is due to 
an offset in the LHC beam position in the transverse plane. The data points taken in the positive and negative halves of the detector along 
$z$ are fitted with sinusoidal curves. }
\label{fig:PixelLeakageCurrent}
\end{figure}

The leakage current in the pixels has also been constantly monitored (figure \ref{fig:PixelLeakageCurrent}).
Annealing took place during a longer shutdown after about 6 fb$^{-1}$ and a shorter technical stop after about 13 fb$^{-1}$ 
delivered integrated luminosity. The LHC collisions are not aligned in the center of the pixel detector
leading to uneven irradiation of the modules along the azimuthal direction, as seen in the leakage current measurement in figure 
\ref{fig:PixelLeakageCurrent} (right).

\begin{figure}[tbp] 
\centering
\includegraphics[width=5cm,height=5cm]{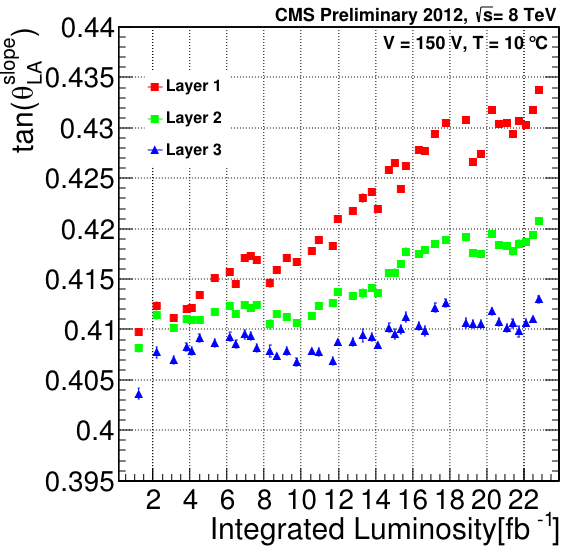}
\includegraphics[width=7cm,height=5.3cm]{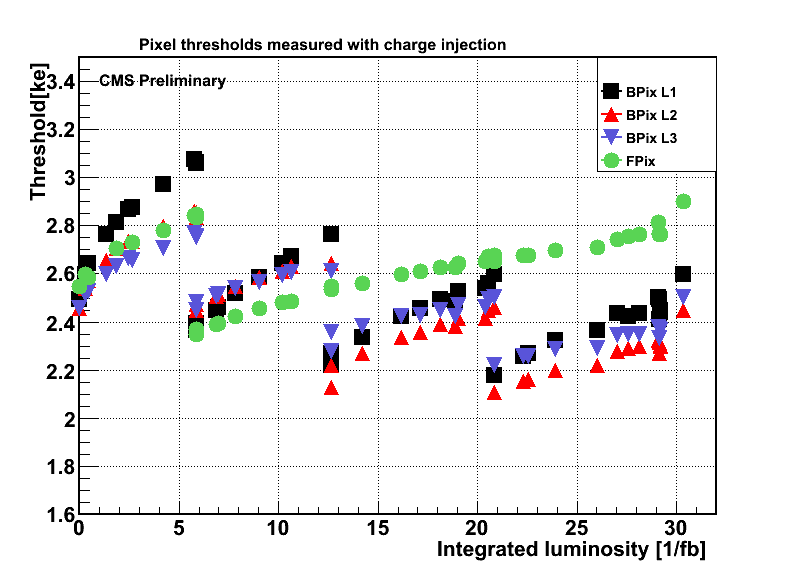}
\caption{The Lorentz-angle (left) and the read-out threshold (right) measured in the pixel detector \cite{TrackerDPG2013}.}
\label{fig:PixelCharge}
\end{figure}

The Lorentz drift of the charges that are generated by high energy particles leads to charge sharing amongst adjacent pixels. 
The angle of the Lorentz drift near the mid-plane was measured repeatedly during 2012 and was found to be increasing with irradiation
(figure \ref{fig:PixelCharge} (left)). 
Charge sharing allows for better resolution as long as the pixels with small signals are nonetheless above read-out 
thresholds. These thresholds are measured in scans by injecting charges into individual pixels in incremental steps. The
thresholds drifted higher with the accumulated irradiation and thus needed to be readjusted multiple times during the LHC 
running period (figure \ref{fig:PixelCharge} (right)).

\begin{figure}[tbp] 
\centering
\includegraphics[width=5cm,height=5cm]{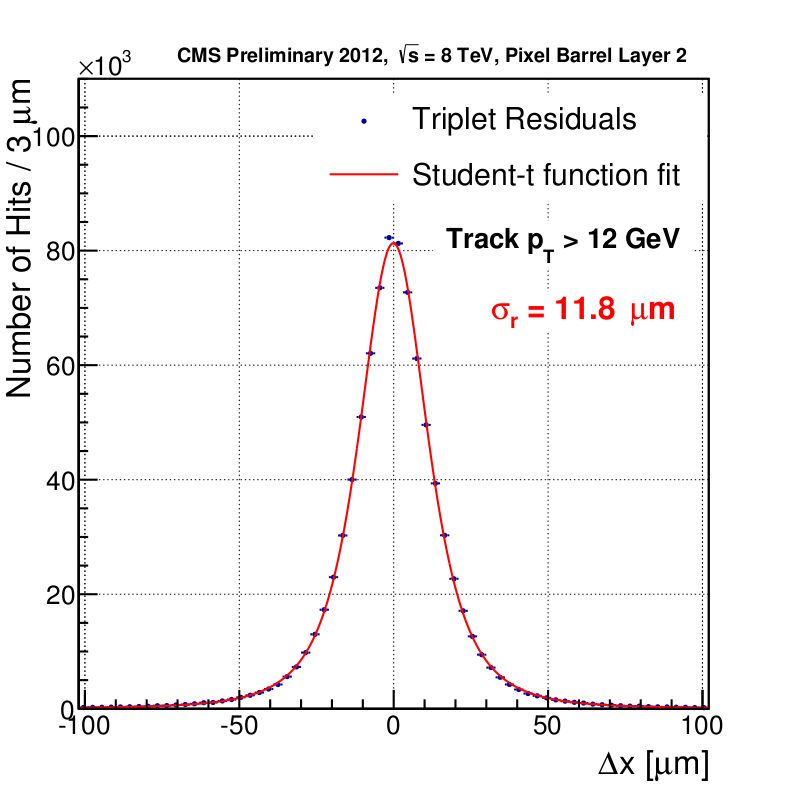}
\includegraphics[width=5cm,height=5cm]{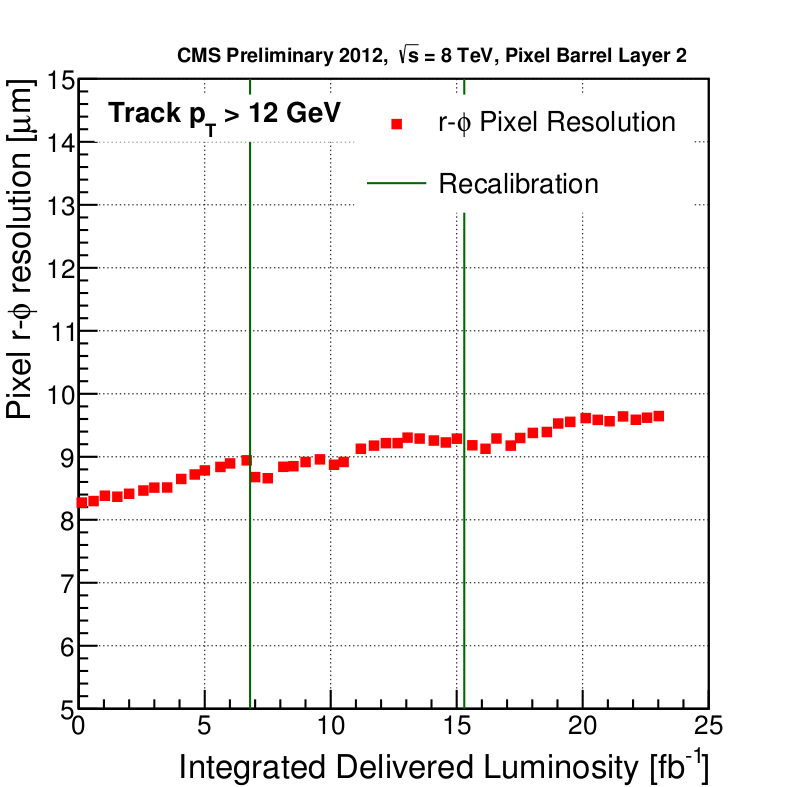}
\caption{Triplet residual distribution for hits in layer 2 of the pixel detector measured in the transverse plane (left) and the evolution of the corresponding 
resolution as a function of delivered integrated luminosity (right).
\cite{TrackerDPG2013}.}
\label{fig:PixelResolutionL2}
\end{figure}

The resolution in the pixel detector is measured by re-fitting hit triplets with their associated tracks whilst omitting from
the fit the hit on the layer under investigation, and by determining the distribution
of the residuals between the expected and measured hit positions. The resolution is the width of the residual distribution after 
removing the errors which propagate from the fit of the other two hits. Figure \ref{fig:PixelResolutionL2} shows a residual distribution
for hits in layer 2 and the evolution of the corresponding resolution as a function of delivered integrated luminosity. The resolution
exhibits a slow degradation over time with two points of improvement corresponding to the times when the pixel read-out thresholds
were recalibrated. The resolution also reflects our understanding of the constantly changing charge gain and Lorentz-angle calibrations.

\section{Alignment of the tracker}\label{sec:TrackerAlignment}

The knowledge of each module's position in three-dimensional space is required for track reconstruction when measurement points localised
on individual modules are placed into the common frame of CMS.
Movements and distortions of the tracker modules are periodically monitored using cosmic ray data and collision tracks by
measuring the change in the distances between the expected trajectory impact points and reconstructed hit positions \cite{CMSTrackerAlignment2013}.

\begin{figure}[tbp] 
\centering
\includegraphics[width=7cm,height=5cm]{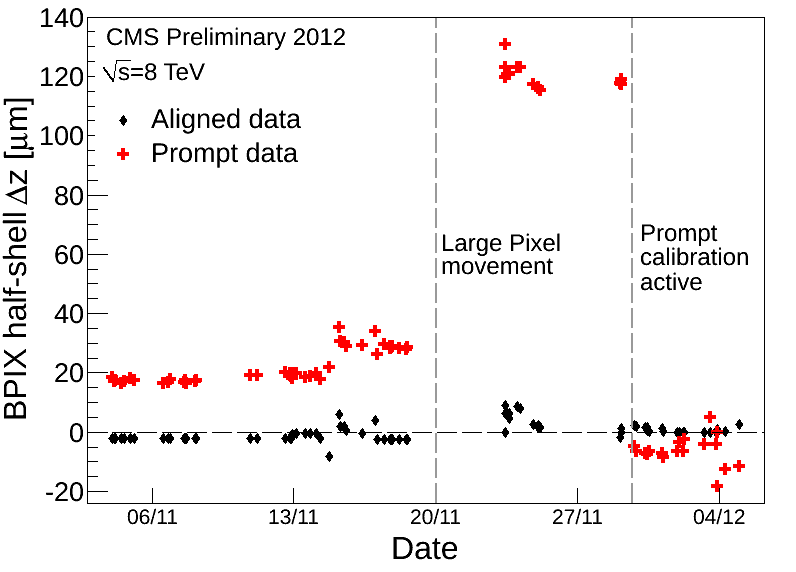}
\includegraphics[width=5cm,height=5cm]{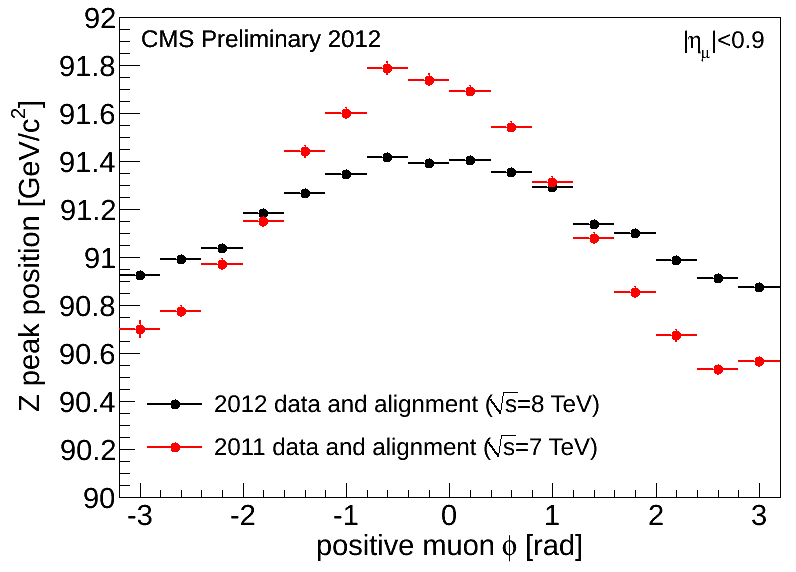}
\caption{Evolution of the longitudinal movements of the pixel-half barrels in time showing results of a thermal cycling on Nov 22 (left) and Z-mass 
bias due to distortions in the transverse plane (right)\cite{CMSTrackerAlignment2013}.}
\label{fig:Alignment}
\end{figure}

Pixel half-barrels are mechanically independent structures that may move with respect to each other in time, especially along the symmetry axis
of the barrel. A relatively large movement with respect to the resolution was recorded on November 22, 2012 in connection with a cooling accident 
(see figure \ref{fig:Alignment}). This movement was detected in a subsequent calibration campaign. Since then, an automatic alignment procedure is 
performed within a few hours of the end of data-taking.

Distortions of the tracker geometry can lead to biases in the reconstructed track curvature. These biases are studied using the reconstructed
invariant mass of the $Z$ boson in $Z\rightarrow\mu\mu$ decays as a function of the positive muon's azimuthal angle (figure \ref{fig:Alignment} (right)). 
The shape of the function implies a distortion of the tracker in the transverse plane, for example, the so-called \emph{sagitta}. Note that 
figure \ref{fig:Alignment} (right) does not represent the CMS muon reconstruction and calibration performance used in physics analyses, rather 
it is the raw tracker input to the event reconstruction.

\section{Track and vertex reconstruction}\label{sec:TrackingVertexing}

The CMS tracking software was designed to operate in an environment with high particle occupancy respecting the limitations on CPU availability. 
It relies on an iterative procedure, each step of which produces a collection of fully reconstructed tracks. Earlier 
sequences search for tracks that are easier to find due to their higher transverse momentum, smaller impact parameter, and larger number of measured 
hits in all tracking layers. Hits associated to these tracks are then ignored in subsequent iterations which, in turn, search for tracks that fulfill 
less stringent requirements and therefore have higher combinatorial complexity, making them more difficult to find.

\begin{figure}[tbp] 
\centering
\includegraphics[height=5cm]{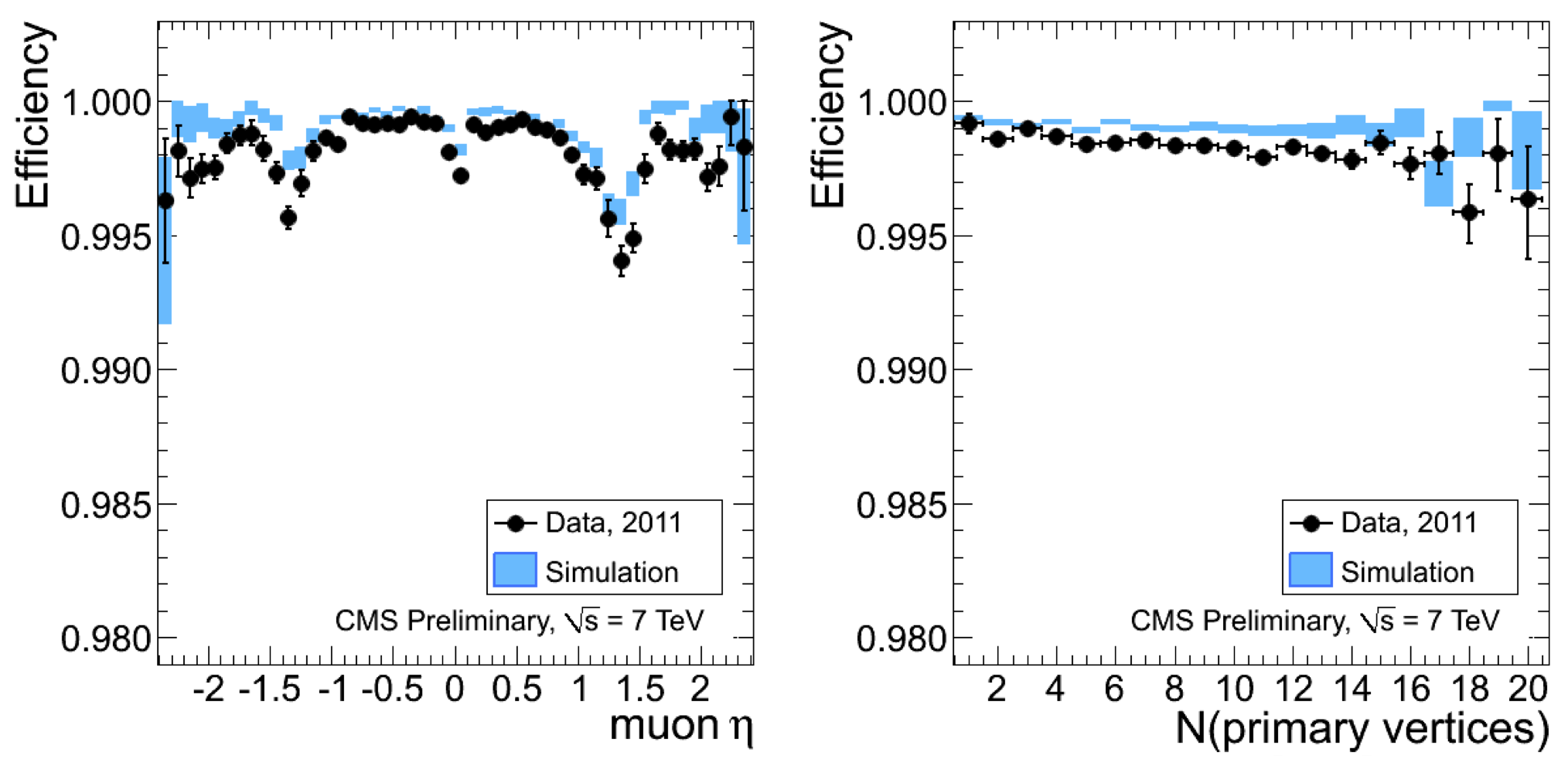}
\caption{Muon reconstruction efficiency in the tracker as functions of pseudorapidity (left) and the number of proton-proton interaction vertices (right) \cite{CMSTracking2012}.}
\label{fig:MuonTnPEff}
\end{figure}

The track reconstruction efficiency was measured with the tag-and-probe method in $Z\rightarrow\mu\mu$ events \cite{CMSTracking2012}. These events 
contain two muons, known as the tag and the probe, which have a combined invariant mass within a window around the nominal mass of the Z boson. The tag muons are
well reconstructed using all information from the tracker and the muon systems. The probe muons are well identified in the muon system only. The tracking
efficiency is then defined as the fraction of the probes which have matching tracks in the tracker. The tracking efficiency as a function of the
pseudorapidity of the tracks and the number of vertices in the event are shown in figure \ref{fig:MuonTnPEff}.

\begin{figure}[tbp] 
\centering
\includegraphics[height=5cm]{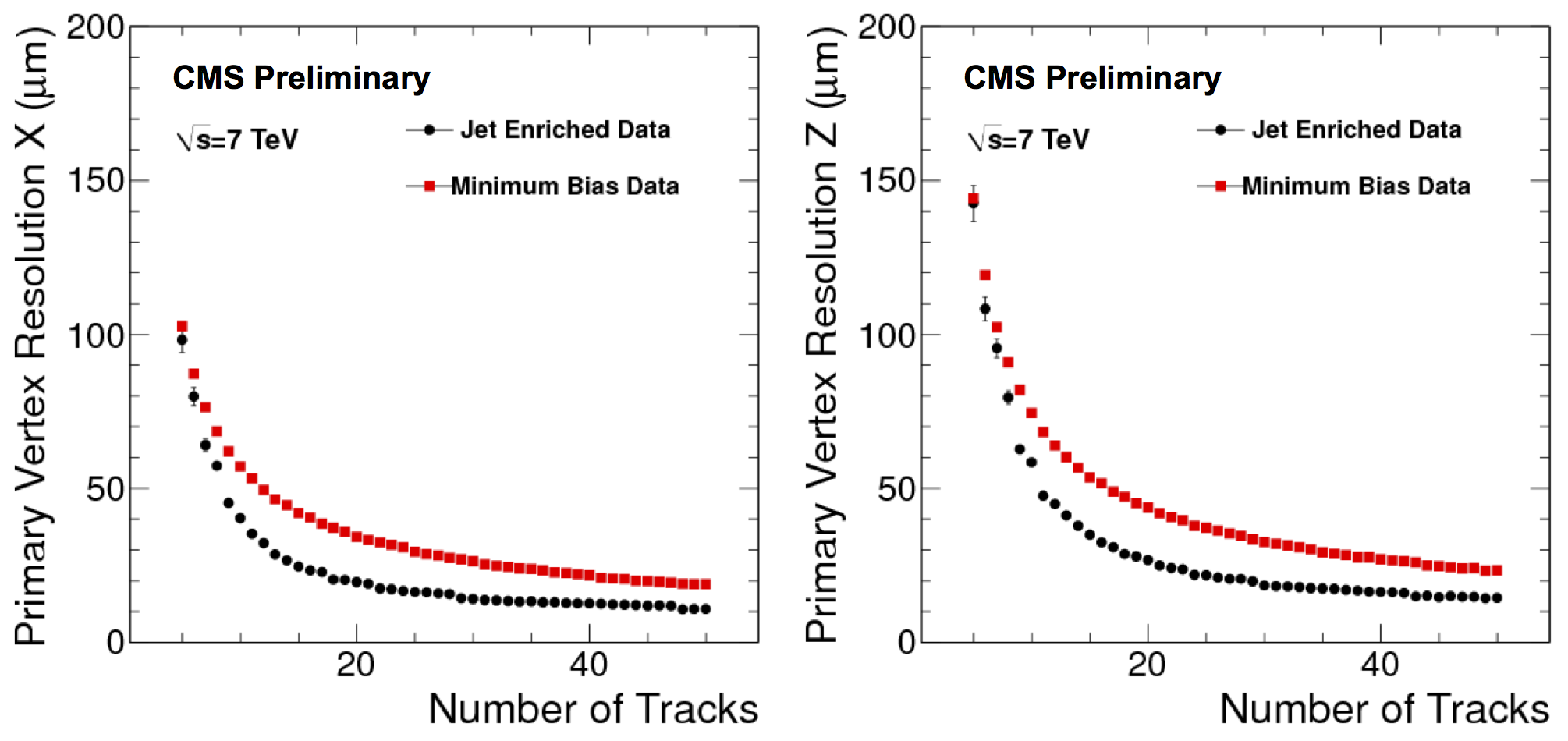}
\caption{Primary vertex resolution in the transverse plane (left) and along the beam-line (right) as functions of the number of tracks 
attached to the vertex \cite{CMSTracking2012}.}
\label{fig:PrimaryVtxRes}
\end{figure}

Each proton-proton interaction point in an event is considered a primary vertex. Its position is measured as the intersection of the associated tracks.
It is located in three steps: first the tracks are identified, then they are grouped according to their vertex 
of origin, and finally the position of each vertex is fitted. The resolution of the vertex position is estimated by the split method \cite{CMSTracking2012} in which the group 
of tracks associated to a given vertex is divided into two sets. The distance between the vertex positions determined from both sets is then proportional 
to the resolution. The vertex resolution in the transverse plane and along the $z$ direction as a function of the number of fitted tracks are shown in figure \ref{fig:PrimaryVtxRes}.

The offline tracking and vertex finding algorithms are also used in the CMS high-level 
trigger (HLT) with configuration parameters customised for fast performance. The HLT processes a 
stream of events at rates up to 100 kHz, and attains output rates up to 400 Hz. Such a large reduction 
in event rate is achieved by utilizing track and vertex information in charged lepton and heavy flavour 
jet reconstruction in order to improve background rejection.

\section{Conclusion}\label{sec:Conclusion and Outlook}

Run 1 of the LHC ended in February 2013. The CMS tracker has been operated successfully
for over three years with excellent performance with regard to detector reliability and tracking. 
During this time, less than 3\% of the detector became inactive and less 
than 5\% of the delivered luminosity was lost due to the tracker. The track
reconstruction methods were able to sustain high pile-up operation with excellent efficiency.

Before the restart of LHC operation in 2014 (Run 2), several improvements are expected to take place.
Some parts of the detector will be replaced and broken modules recovered such that the
fraction of inactive modules is expected to decrease below 1\%. In order to compensate
for the accumulated radiation damage, the coolant temperature will be lowered in both
the pixel and strip detectors. Improved hit reconstruction will take into account the
radiation induced changes in the sensors. Centering the pixel detector around the LHC beams
will result in more uniform distribution of radiation damage along the azimuthal angle.
In Run 2, both collision energy and instantaneous luminosity will nearly double. 
The tracking algorithms are going to be readjusted in order to meet the challenges presented by
the higher occupancy.

\end{document}